\documentclass[preprint,12pt]{elsarticle}

\usepackage{graphicx}
\usepackage{hyperref}
\usepackage{amssymb}

\usepackage{lineno}

\journal{Journal Name}

\begin{document}

\begin{frontmatter}


\title{Mobility-based prediction of SARS-CoV-2 spreading}



\author[1,2,3]{Lorenzo Chicchi}
\author[1,2,3]{Lorenzo Giambagli}
\author[1,2]{Lorenzo Buffoni}
\author[1,2,3]{Duccio Fanelli}

\address[1]{Department of Physics and Astronomy, University of Florence, Sesto Fiorentino, Florence, Italy}
\address[2]{CSDC, University of Florence, Sesto Fiorentino, Florence, Italy}
\address[3]{INFN Sezione di Firenze, Sesto Fiorentino, Florence, Italy}

\begin{abstract}
The rapid spreading of SARS-CoV-2 and its dramatic consequences, are forcing policymakers to take strict measures in order to keep the population safe. At the same time, societal and economical interactions are to be safeguarded. A wide spectrum of containment measures have been hence devised and implemented, in different countries and at different stages of the pandemic evolution. Mobility towards workplace or retails, public transit usage and permanence in residential areas constitute reliable tools to indirectly photograph the actual grade of the imposed containment protocols. In this paper, taking Italy as an example, we will develop and test a deep learning model which can forecast various spreading scenarios based on different mobility indices, at a regional level. We will show that containment measures contribute to ``flatten the curve'' and quantify the minimum time frame necessary for the imposed restrictions to result in a perceptible impact, depending on their associated grade. 
\end{abstract}


\end{frontmatter}


\section{Introduction}
\label{S:1}

Machine Learning (ML) \cite{bishop_pattern_2011, hastie2009elements} has been extensively employed in the context of time series modeling and forecasting \cite{fawaz2019deep}. Groundbreaking applications in natural language processing \cite{chowdhary2020natural}, financial forecasting \cite{tino2001financial}, speech recognition \cite{deng2013new} have earned this particular subfield of ML lots of investments and attention. Notably, the use of Deep Neural Networks \cite{lecun2015deep, Goodfellow-et-al-2016}, with respect to traditional approach to time series analysis, enabled the algorithm itself to learn from the data the relevant variables and their associated correlations. Following the rapid spreading of SARS-CoV-2, numerous attempts have been made for predicting the time evolution of epidemics across different spatial scales \cite{fanelli2020analysis, carletti2020covid,vespignani2020modelling}. To this end ML techniques have been also employed \cite{shastri2020time,farooq2020deep,mathew2020deep}. Although very accurate and useful, these models often lack the ability to incorporate the effects of containment measures as implemented by local governments and solely rely on selected epidemiological variables (e.g. number of tests performed, number of deaths) to predict the spreading of the virus. The putative impact of different containment strategies as devised by local governments is hence customarily modeled by resorting to standard epidemiological tools \cite{kraemer2020effect,ilin2020public},a choice which potentially limits the predictive ability of the trained ML devices. Starting from these premises, we suggest that mobility indices provide solid, almost real-time, indicators of the implemented containment strategies. When included in the training, they are processed as key information for future forecasting of ML algorithm. A self-consistent argument allows in turn to estimate the time it takes for the imposed mobility restrictions to materialize in an effective drop of the curve of infected individuals.

In the following, we will describe the adopted machine learning approach which is tailored to predicting the SARS-CoV-2 epidemic evolution in the twenty regions of Italy \footnote{\footnotesize Valle d'Aosta, Piemonte, Lombardia, Trentino - Alto Adige, Veneto, Friuli Venezia Giulia, Liguria, Emilia Romagna, Toscana, Marche, Umbria Lazio,Abruzzo, Molise, Campania, Apulia, Basilicata, Calabria, Sicilia e Sardegna}. The model is trained by using the time series of selected epidemic quantities (number of infections, number of death, etc..) and includes information on the population mobility.
We will show that, by looking at epidemic and mobility trends during the $n_p$ past days, the model is able to return sensible information on the values of a target epidemiological parameter in the next $n_f$ days. Working in the proposed framework, we are also  able to estimate the time needed for the imposed restriction to yield consequences that can be appreciated at the scale of the whole community in terms of reduction of hospitalized individuals. To this end, we consider different grades of imposed restrictions on individual mobility ranging from a complete, nationwide lockdown to milder, regional-level restrictions to virtually no restrictions at all.

\section{Methods}
\subsection{Architecture}
We worked with Recurrent Neural Networks (RNN) \cite{Goldberg}, a class of deep learning architectures widely used in time series machine learning modeling. Such architecture is designed to be sensitive to the ordering of the elements in the input sequence \cite{Goldberg}. This is achieved by introducing an inner state vector that is updated by the network itself, during each successive iteration. This latter vector allows the network to ``keep memory'' of the past input values. RNNs suffer of the so-called vanishing gradient problem: the gradients in later steps of the sequence fade away quickly in the backpropagation process, without reaching earlier input signals and thus making it hard for the RNN to apprehend and correctly incorporate long-range dependencies \cite{Goldberg}. To oppose this problem, gating-based architectures, such as the Long short-term memory (LSTM), have been proposed \cite{hochreiter1997long}. Trainable vectors, called gates, are accommodated  for in the architecture and control the inner state update, at each iteration. This technical solution makes it possible for the network to ``forget" or ``store" the novel bits of information that are processed at each time step, along the sequence of collected events. In this way, early information deemed crucial for handling the forecasting task can be stored in the bulk while, recent inputs, identified as unessential, are safetly removed from the memory kernel. This is precisely the reason why we have decided to employ a LSTM-like architecture for the problem at hand. In the following we shall operate with a deep architecture composed by two LSTM hidden layers of $300$ and $20$ nodes, respectively. Moreover, an additional dense layer is introduced to produce the sought output. Further, use is made of Adam \cite{kingma2014adam} optimizer with a learning rate of $0.0005$. The batch size is set to $100$ and the number of epochs during the learning procedure is assumed equal to $100$. We hand picked these hyper-parameters without any ad-hoc optimization.

\subsection{Data set}
\begin{figure}[h!]
    \centering
    \includegraphics[width=\textwidth]{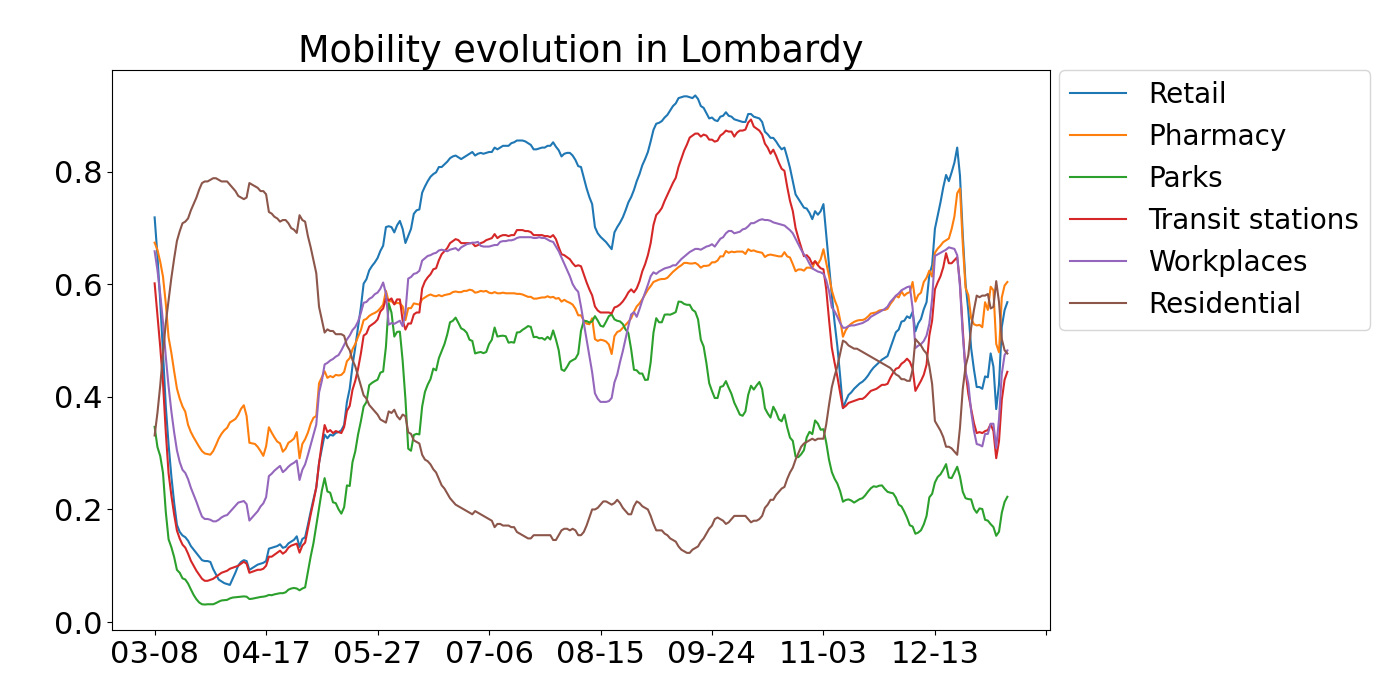}
    \caption{Time evolution of mobility parameters in Lombardy. A seven-days moving average was performed to filter out the weekly fluctuations and highlight the global trend. }
    \label{Data1}
\end{figure}

\begin{figure}[h!]
    \centering
    \includegraphics[width=\textwidth]{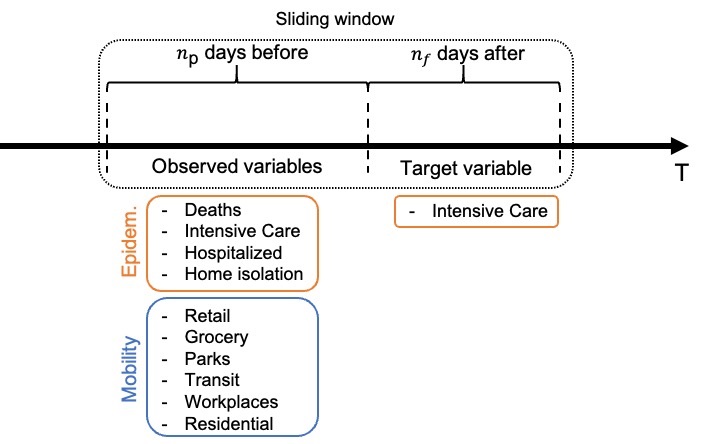}
    \caption{Each data consist of an ensemble of input parameters (concerning both epidemiological and mobility quantities and associated to the $n_p$ days that precede the day where the analysis is carried out.) and an output set that contains our forecast. This is the number of occupied Intensive Care (IC) units in the following $n_f$ days (from the last day of observation).}
    \label{Scheme}
\end{figure}

The dataset consists in discrete daily series of length $T$ of selected epidemic and mobility parameters for each of the 20 regions in Italy. More specifically, we focus on the following quantities:  (i) number of patients in \textit{intensive care } (ii) number of  \textit{hospitalized} patients (iii) number of patients in \textit{home isolation} (iv) number of \textit{deaths}. Data from the COVID-19 Community Mobility Reports of Google \cite{Google} are employed to track the change in time of the degree of mobility, as associated to different regions of Italy. We calculate in particular the evolution as percentile change from baseline values\footnote{\footnotesize The base value is defined as the average value in the five weeks between 3th January and 6th February 2020 for the considered week day, as explained in \cite{Google}} of the reported mobility indexes in the following areas:
\begin{enumerate}
    \item retail and recreation
    \item grocery and pharmacy
    \item parks 
    \item transit stations
    \item workplaces 
    \item residential
\label{ParMob}
\end{enumerate}
In Fig. \ref{Data1} the evolution of the reference mobility indicators are displayed for the case of Lombardy. 
The impact of the imposed restrictions on the mobility indexes can be clearly appreciated by visual inspection of the depicted global trends. It is hence surmised that the aforementioned mobility indicators provide a faithful barometer to gauge the actual impact of the imposed containment measures. As such, they could be accounted for when training the LSTM to forecasting the future evolution of the epidemics.\\
Combining all these together, for each day of the time series, we access 10 distinct parameters, either referred to the evolution of the epidemics or the mobility trends. Input entries are normalized so as to operate, during training stage, with quantities that span a definite range (details are provided in the Figures' captions). The above information are used in the supervised learning problem and organized as follows. The input vector collects information referred to the past $n_p$ days. Epidemics and mobility data sum up to a total of 10 scalar parameters per day of acquisition. The output target vector has length $n_f$, the time horizon of the prediction. More specifically, each entry of the output vector returns a prediction for the number of patients in intensive care (IC) units, at the day of forecast, up to $n_f$ days in the future. 
A schematic representation of data structure and processing handling is provided in Fig. \ref{Scheme}. The mobile window is made to slide along the scrutinised time series, day after day. For each position of the window, the information stemming from the $n_p$ preceding days (including the current day of observation) are acquired and confronted with the desired output, the number of occupied IC units in the future $n_f$ days. During the training phase, this information is used to adjust the weights of the LSTM. When properly trained, this device is used for forecasting purposes by letting the sliding window to explore a portion of the times series not supplied during the learning phase. The computing apparatus is fed with the needed input information referred to the past $n_p$ days (including the day of elaboration) to anticipate the future (the following $n_f$ days) in terms of expected COVID-19 patients necessitating IC units. 

Summing up, the training data set is made of $20 (T-n_p-n_f)$ examples (that is, couples \textit{input-target}) where the factor 20 stems from the number of considered regions. In the analysis reported below $n_p=21$ (meaning that we process data from the last 21 days of observation) and $n_f=7$ (hence, for each position of the sliding window, we look forward in time with a horizon of prediction that covers one week in the future).

The data set is divided into two subsets: \textit{training and test sets}. The first set is used to train the model, whereas the second one is employed to test the ability of the trained device to cope with data that were not supplied during the learning stages. To probe the robustness of the method we have devised two different procedures to split the data in training and test set, respectively. These are listed in the following:.
\begin{enumerate}
\item \textit{The training set consists in a limited segment of the available times series}. In this second case, the training procedure is carried out by solely 
    employing data up a prescribed date. The future evolution of the system, beyond the last day of the processed observation, is used  to test the model. This makes it possible to test the performance of the LSTM model against data that refer to a time window not processed during learning.
    \item \textit{The training set is a sub-set of the available regions}. In this case, the learning process is carried out over the entire length of the time series associated to a subset of the 20 regions. The accuracy of the prediction is tested against data referred to regions that were not used during the learning phases.
\end{enumerate}
In the next subsection we will discuss the obtained results and validate the model as a viable tool to anticipate SARS-CoV-2 spreading across the Country.

\section{Results}

\begin{figure}[h!]
    \centering
    \includegraphics[width=1\textwidth]{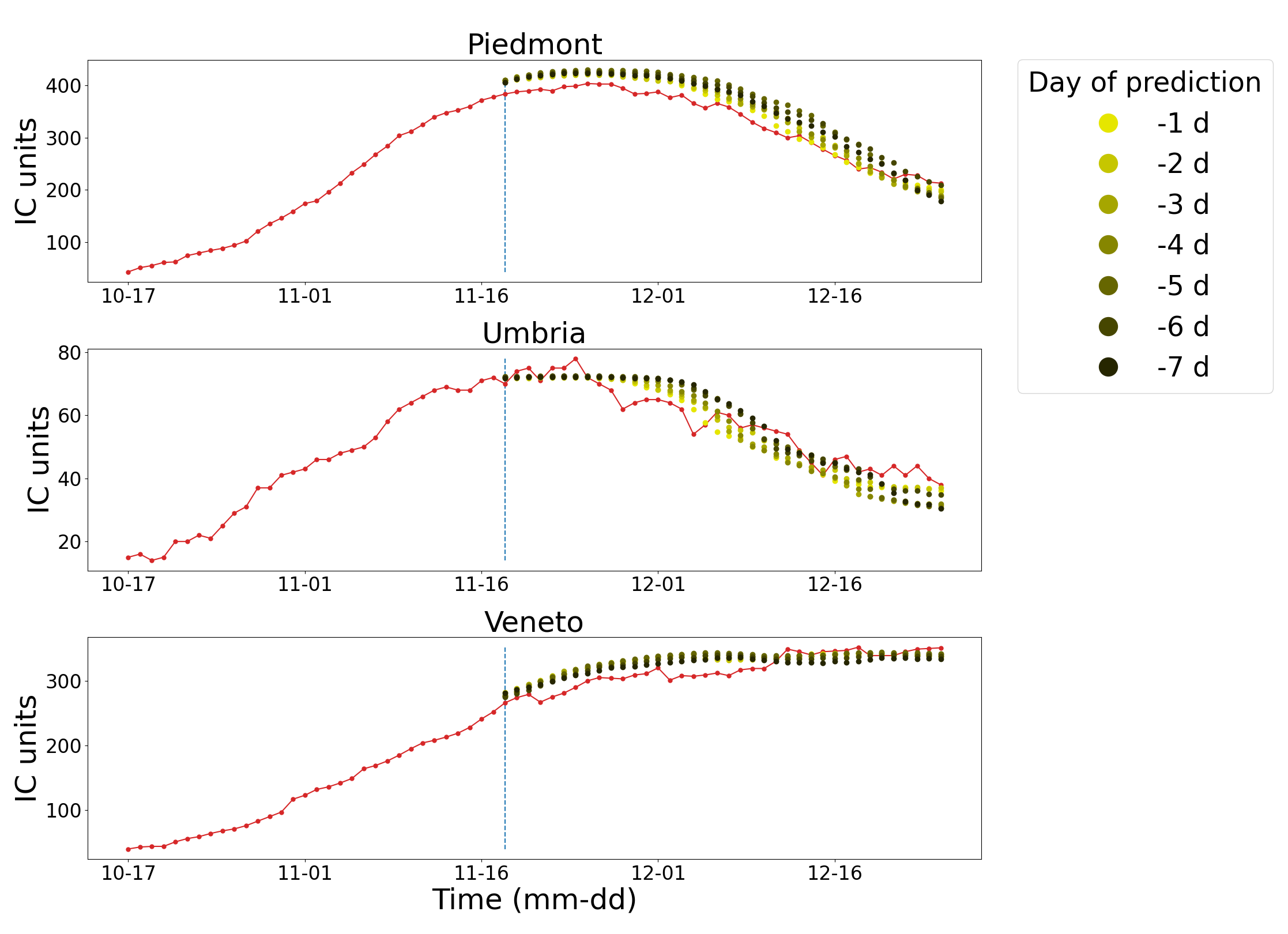}
    \caption{Predicted evolution as compared to the experimentally recorded time series: the plotted curves refer to the number of occupied Intensive Care  (IC) units  by SARS-CoV-2 patients in Piemonte, Umbria and Veneto. Red lines stand for the observed (hence, real) evolution. Coloured dots represent the forecast of the LSTM model.  Yellow dots are predictions that look at one day in the future. Black symbols rely instead on processing one week old  data. Different color gradings, ranging from yellow to black, interpolate between these two limiting scenarios. In this case $n_p$ is set to 21. Data are rescaled by using, for each variable of the set, its corresponding maximum value, as displayed in the training interval. This latter value is also used to normalize data from the test set, so that only information from the training set are effectively employed. }
    \label{PanA}
\end{figure}

We used the architecture and data set as described in the previous section to define a learning problem that allows one to predict the evolution of the number of intensive care units (IC) occupied by COVID-19 patients in different regions of Italy.\\  
We begin by adopting the first of the two aforementioned frameworks. This implies dealing with the full set of available time series, up to a given time, for the training phase. The trained network is then employed to forecast the evolution of the epidemics.  Results are reported in Fig. \ref{PanA} for a subsets of regions, namely Piedmont, Umbria and Veneto. The evolution of occupied IC units (orange trace) is nicely predicted by the model (coloured dots). 

For each day, the number of truly occupied IC units is compared to the corresponding value, as predicted by the LSTM with different time horizons. More specifically, yellow dots refer to predictions which exploit information made accessible up to the preceding day. On the opposite limit, black dots are forecast that process information older than one week (7 days).  Intermediate color grades refer to predictions which interpolate between these two extremes.  The data reported (Fig. \ref{PanA}) are obtained by training the LSTM with data up to November 16th, where the dashed line is positioned. From here on, predictions are obtained by sliding the computing window (as depicted in Fig. \ref{Scheme}) forward in time. The information relative to the $n_p$ input days are processed and used to anticipate the expected load of IC units in the next $n_f$ days. The forecasted evolution agrees pretty well with the observed curve of occupied IC units. Remarkably, the position of the peak is nicely captured by the computed time series which is hence capable to anticipating the evolution of the examined system. 

\begin{figure}[h!]
    \centering
    \includegraphics[width=\textwidth]{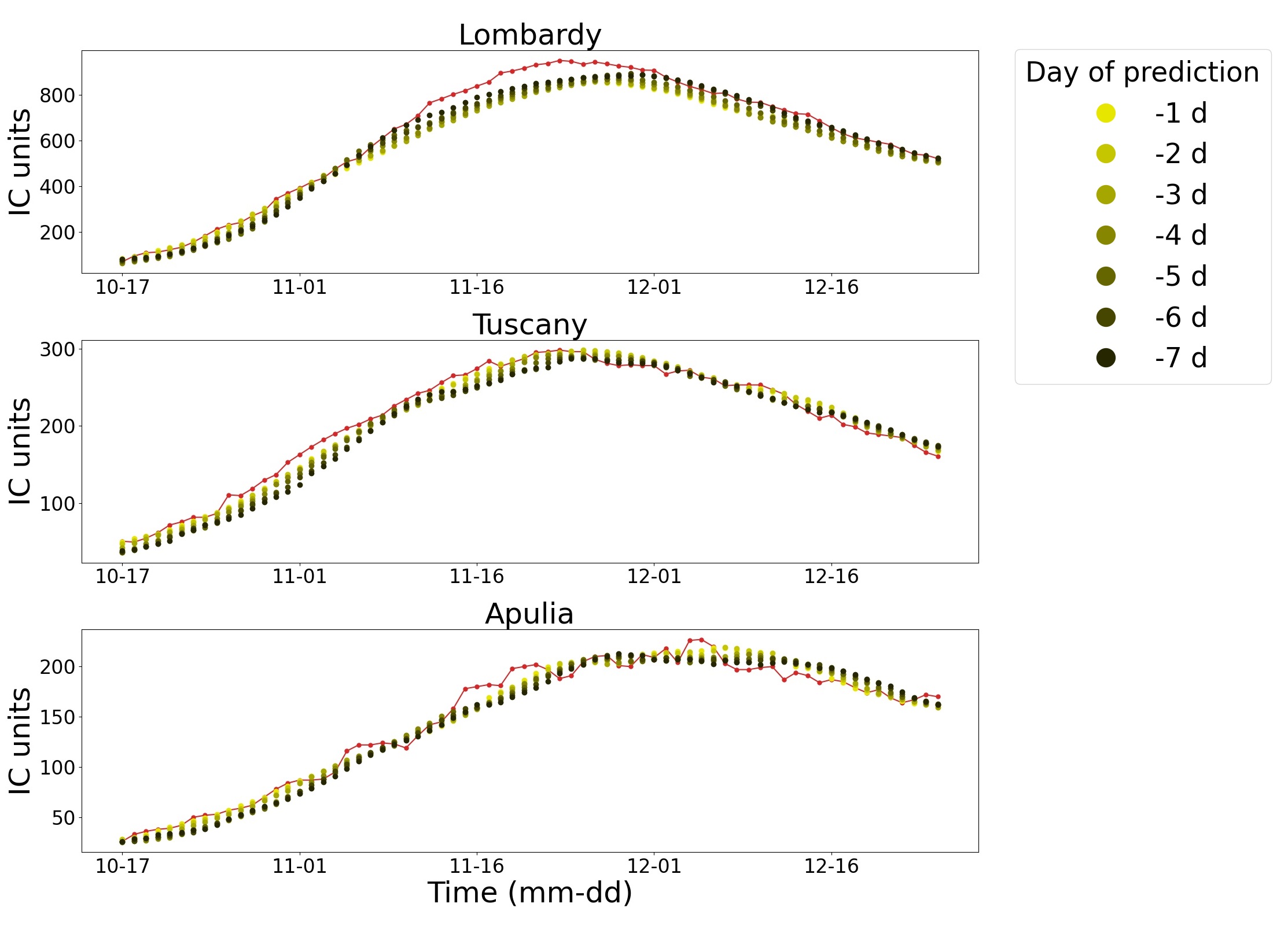}
    \caption{The number of occupied IC is plotted against time, expressed in days. Red traces reflect the observed, hence true, numbers. Colored dots stand for the LSTM forecasst. Here, the analysis is carried out for three regions that were not part of the training set. The adopted color code is specified in the caption of Fig. \ref{PanA}. The analysis is carried out by using $n_p=21$. Here, epidemiological data are normalized by an arbitrary constant that we assume to extensively scale with the size of the examined region.}
    \label{PanB}
\end{figure}

In Fig. \ref{PanB} the results obtained when dealing with the alternative setting as listed above, are depicted. As mentioned, we now train the model by focusing on a subset of the available regions and use this knowledge to predict the evolution of IC units occupied by COVID-19 patients in regions that were not supplied as part of the training set. Color are assigned following the same code introduced above: yellow dots refer to prediction that looks to just one day in the future. Black dots stand for the opposite extreme: the LSTM anticipates the evolution one week ahead in time. The agreement between predicted and observed times series is again remarkable.

\begin{figure}[h!]
    \centering
    \includegraphics[width=\textwidth]{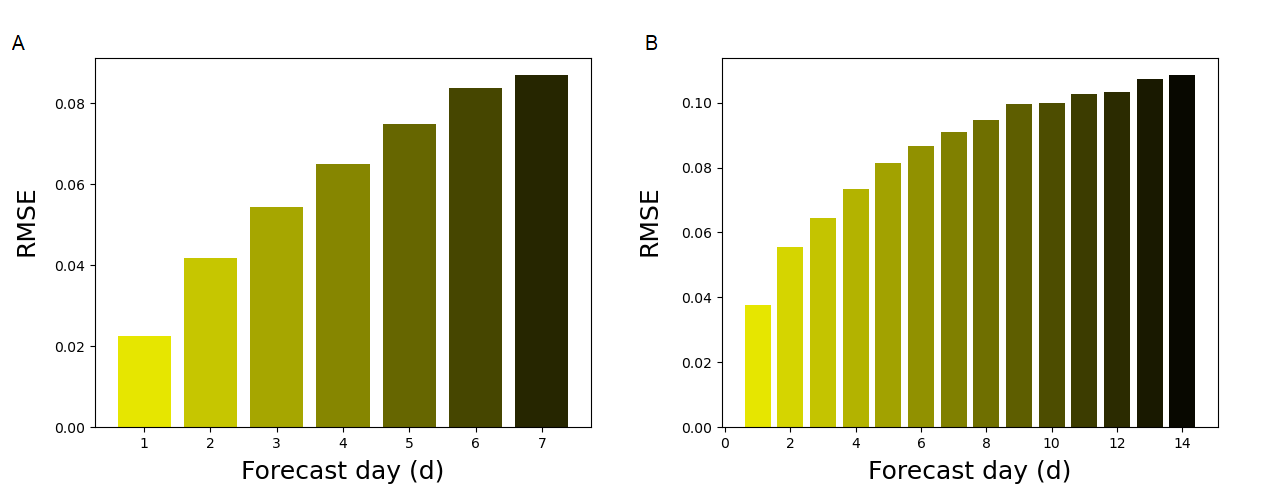}
    \caption{Root-Mean-Square Error (RMSE) computed on the test set: the error compares the real evolution of IC occupation number and the expected evolution on the basis of the LSTM prediction at day   $d\in[1,n_f]$, from the last processed observation. Panel (A) and panel (B) referred to LSTM models with $n_f=7$ and $n_f=14$, respectively. }
    \label{PanC}
\end{figure}

In Fig. \ref{PanC} the Root-Mean-Square Error (RMSE) associated to the predictions is plotted.  Each bar represents the error made when trying to predict the target values $d \in [1, n_f]$  days in the future, where the parameter $n_f$ defines the forecast horizon of the model. The RMSE is computed over the test set. Panel A of Fig. \ref{PanC} is referred to $n_f=7$, whereas panel B is obtained for $n_f=14$. As  expected, the accuracy goes down when $d$ becomes larger. Although the model with larger $n_f$ allows us to make early predictions, the accuracy of the predictions get worse when confronted with actual data: a lower accuracy is found not only for distant predictions but also for closer ones. The choice $n_f=7$ is a compromise between the need to cope with reliable predictions, on the one side, and the request of  imposing a plausible temporal horizon, i.e. useful for forecast, on the other.

In the following, we will shortly elaborate on the role of the mobility and highlight its reflexes on the evolution of the epidemics.


\subsection{Change the mobility}

\begin{figure}[h!]
    \centering
    \includegraphics[width=\textwidth]{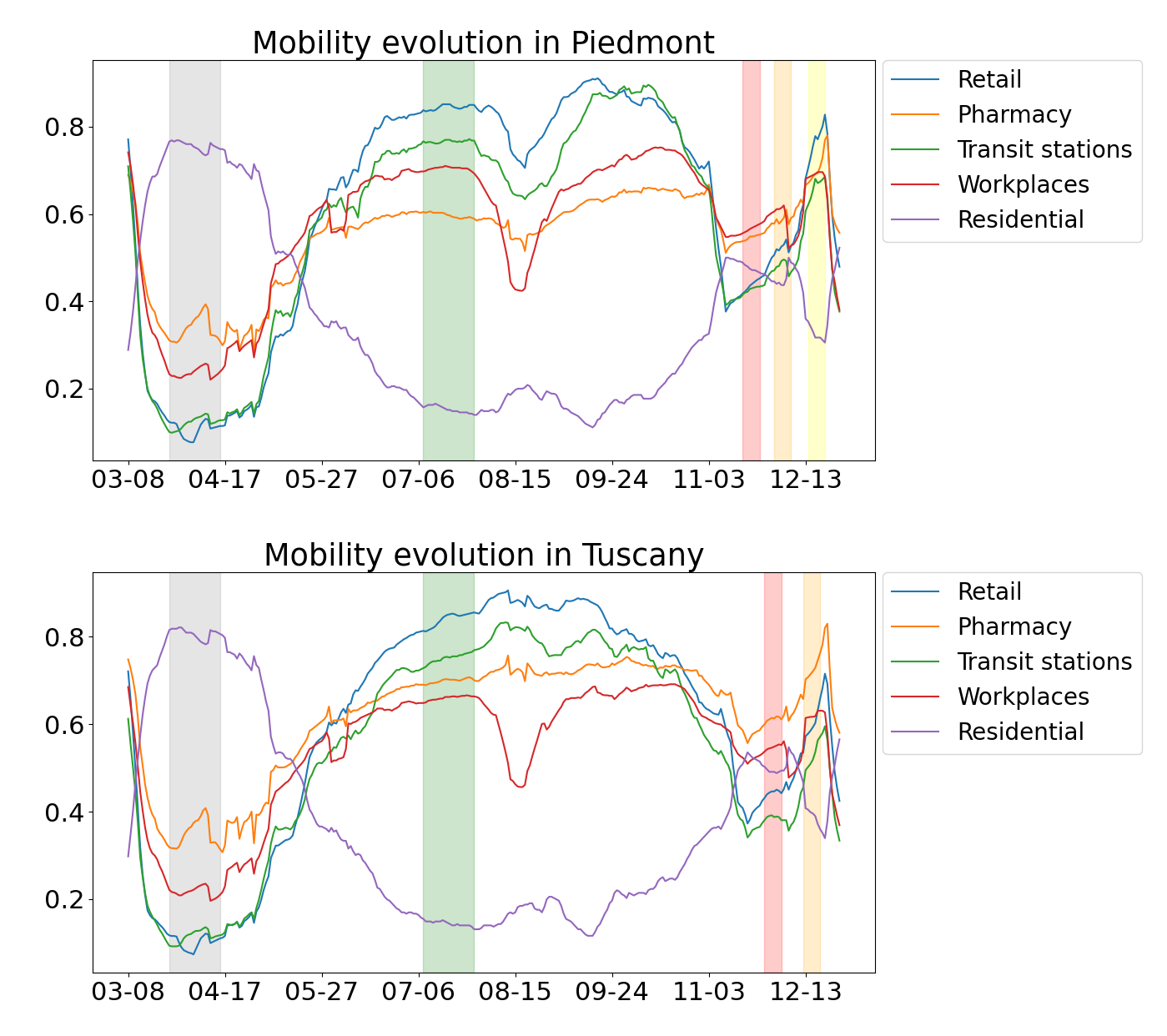}
    \caption{Mobility evolution in Piedmont and Tuscany, with reference to five distinct categories, as outlined in the annexed legend. Data are from \cite{Google} and have been normalized to yield quantities that range in the interval $[0,1]$. A moving average of 7 days is operated so as to remove spurious weekly fluctuations. The four vertical bars define the time intervals over which the reference mobility values have been estimated.}
    \label{Mob}
\end{figure}

Different containment measures have been imposed to contrast the spreading of the COVID-19 epidemic. Such measures, like social distancing and lockdown, result in a clear impact on the mobility. In Fig. \ref{Mob} the evolution of five normalized mobility parameters are plotted for Piedmont and Tuscany. Data are processed by operating a 7 days moving average to obtain smoother profiles and remove  weekly fluctuations. The averaged mobility indexes display global trends which bear the imprint of the containment measures as imposed by national and local authorities. To support this claim, for each region, five time intervals associated to different containment measures have been identified. At the beginning of the time series, a depression of all the mobility scores is detected (except for the parameter that quantifies mobility in residential areas -- purple lines -- which in general, and as expected, shows opposite trends as compared to those stemming from other parameters). This has to be put in relation with the strict lockdown taken by the Italian government in the spring of 2020. Subsequently, the curves associated to the parameters of not-residential areas grow up until they reach a new plateau. The plateau follows a no-restrictions (or few-restrictions) period, during the summer, when containment measures had been relaxed. Other characteristic periods can be indeed identified, specifically at the end of November and at the beginning of December. This last segment of the recorded time series is indirectly influenced by the 
introduced color code labelling of the regions, as reflecting the degree of local severity of the epidemics.
Each region is in fact associated to a color (respectively, yellow, orange and red in ascending order of severity) and a different level of restrictions are adopted depending on the region color. The correlation between the actual severity of the imposed restrictions and the displayed mobility trends can be clearly appreciated by visual inspection of Fig. \ref{Mob}. To help visualization few (colored) vertical stripes are depicted which refer to different conditions of the mobility, as outlined above. The first bar, colored in grey, is traced in correspondence of the strict lockdown back in the spring 2020. By averaging over the selected time interval (the width of the greyish bar) we obtain an average estimate of the mobility parameters, as associated to the lockdown phase. Similarly, the other depicted bars identify other characteristic instances of the epidemics evolution: the green stripe is meant to select mobility score referred to the summer 2020. The red/orange/yellow bars identify the status of the region, as follow the novel strategy to label the severity of the disease at the local scale. Also in this case, by averaging over the width of the corresponding intervals, one obtains a set of values for the mobility indexes which indirectly reflect the imposed containment action (from draconian lockdown to no restrictions, via the intermediate settings as associated to different labelling colors).

\begin{figure}[h!]
    \centering
    \includegraphics[width=\textwidth]{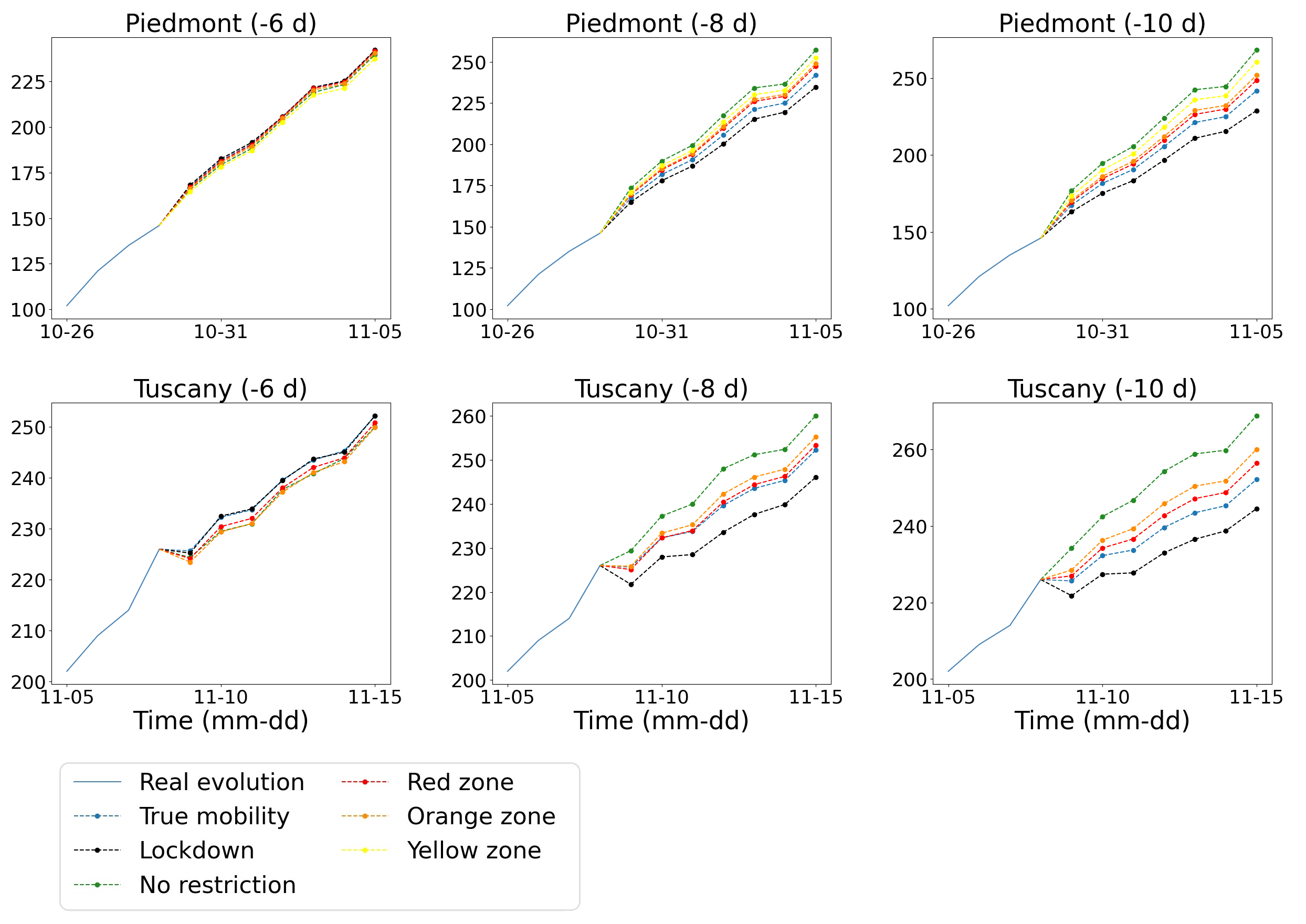}
    \caption{Predictions of IC units occupied by COVID-19 patients in Piedmont and Tuscany made under the hypotheses of 6 different past mobility scenarios: true mobility (blue dots), lockdown (black dots), no restrictions (green dots), red zone (red dots), orange zone (orange dots) and yellow zone (yellow dots). In each column, from left to right, the change in the mobility scores is operated at a different day, as measured from the time the first prediction is made (see annexed legend).}
    \label{PanD}
\end{figure}

This information can be used in the attempt to predict the role of an enforced modulation of the mobility, as follow the different scenarios recalled above. More specifically, at any given day, one can change the mobility entries as supplied to the trained LSTM (by fishing from the aforementioned alternative classes, identified via the corresponding averaged entries). The aim is  examine the ensuing effect which materializes at the level of the forecasted evolution of the occupied IC units, the target of the LSTM.  In Fig. \ref{PanD} the result of the analysis is displayed for two reference regions, although the reached conclusion holds in general. A punctual modulation of the mobility (i.e. a change in the mobility that is confined to just one day) produced sensible changes in the predicted hospitalization, the response being more marked the stricter the reduction of the mobility being imposed. Remarkably, and according to the LSTM, the effect of a local change in the mobility becomes visible $8$-$10$ days in the future, a plausible outcome of the analysis which calls for a timely planning of the containment protocols. On the basis of the above, it is hence surmised that machine learning schemes of the type here analyzed could help devising  optimal strategies for an intelligent combination of openings and closures, at the local scale. Furthermore, notice that the lag time quantified above provides an a posteriori justification for choosing $n_f=7$ as a forecast horizon of the LSTM machinery.

\section{Conclusions}

To summarize our findings, using a simple LSTM model trained on both epidemiological and mobility data we were able to correctly forecast the spreading of SARS-CoV-2 across different regions and at different times. Our model proved robust to alternative train/test splits in the spatial (hold out a region) and temporal (hold out a temporal interval) domains. The choice of employing available information on human mobility  constitutes the novelty of the proposed approach. The obtained forecasts are indeed shown to sensibly depend on the imposed mobility scores. When artificially reducing the degree of imposed mobility, yields a consistent flattening of the curve of expected occupied intensive care units. Importantly, the effect of an abatement of the mobility materializes in a consequent contraction of the occupied IC. The contraction becomes visible after $8$-$10$ days, from the time the mobility change became effective. Interestingly, punctual mobility stops (one day) seem to generate a noticeable effect on the predicted IC occupation curve. Elaborating further along these lines could help devising viable strategies to oppose the spreading of the epidemics, with a minimal impact on both social and economical activities.






\bibliographystyle{unsrt}
\bibliography{sample.bib}







\end{document}